\documentclass[prl,aps,twocolumn,showpacs]{revtex4-1}

\usepackage{amsmath,amssymb,amstext,amsthm,amscd,mathrsfs,eucal,bm,graphics,graphicx,overpic}

	\def\beq{\begin{equation}}
	\def\eeq{\end{equation}}
	\def\eref#1{(\ref{eqn:#1})}
	\def\elab#1{\label{eqn:#1}}
	
	\def\fref#1{\ref{fig:#1}}
	\def\flab#1{\label{fig:#1}}

	\def\nor#1{\lVert #1 \rVert}

	\def\Pe{\text{Pe}}
    
	\def\Da{\text{Da}}

	\def\bu{\boldsymbol{u}}

	\def\G{\mathscr{G}}

	\def\bphi{\boldsymbol{\phi}}

     \def\bphi{\boldsymbol{\phi}}
   \def\b0{\boldsymbol{0}}

\def\strutdepth{\dp\strutbox}
\def\nw#1{\strut\vadjust{\kern-\strutdepth\vtop to0pt{\vss\hbox to\hsize {\hskip\hsize\hskip5pt$\leftarrow$\hss\strut}}}{\em #1}}
       
\begin{document}

\title{\bf Front propagation in cellular flows for fast reaction and small diffusivity}
\author{Alexandra Tzella${}^{1}$ and Jacques Vanneste${}^{2}$}
\affiliation {${}^1$School of Mathematics,
University of Birmingham, United Kingdom \\
${}^2$School of Mathematics and Maxwell Institute for Mathematical Sciences, University of Edinburgh, United Kingdom}
\date{\today}
\begin{abstract}
 	We investigate the influence of fluid flows on the propagation of chemical fronts  arising in FKPP 	type models. 
	We  develop an asymptotic theory for the front speed in a cellular flow in the limit of
	small  molecular diffusivity and fast reaction, i.e., large P\'eclet ($\Pe$) and Damk\"ohler ($\Da$) numbers.
    The front speed is expressed in terms  
	 of a periodic path -- an instanton -- that 
	minimizes a certain functional. This  leads to an efficient procedure to calculate the front speed, and to closed-form 
	expressions for $(\log\Pe)^{-1}\ll\Da\ll\Pe$ and for $\Da\gg\Pe$.
	Our theoretical predictions are compared with (i) numerical solutions of an eigenvalue problem and (ii)  simulations of the advection--diffusion--reaction equation.
\end{abstract}
\maketitle

The spreading of chemical or biological populations in fluid flows 
is a fundamental problem  in many areas of science and engineering with applications ranging from plankton blooms   to 
 combustion  
\cite{Tel_etal2005,NeufeldHernandezGarcia2009}. 
In the absence of flow, this spreading  results from the competition between spatial diffusion, local growth and saturation, and leads to the formation of wave fronts that travel undeformed at constant speed \cite{Murray}. 
A number of theoretical results describe the influence that   divergence-free, spatially smooth flows have on  such fronts 
(see   \cite{Xin2000,Xin2000b} for comprehensive reviews). These have stimulated   experiments using a variety of  reactions and flow configurations   
\cite{Leconte_etal2003,Paoletti_etal2006,SchwartzSolomon2008}, with much effort devoted to steady spatially periodic cellular flows \cite{PaolettiSolomon2005steady,PocheauHarambat2006,PocheauHarambat2008,Thompson_etal2010}.  

Using the classic model of Fisher \cite{Fisher1937} and Kolmogorov \emph{et al}. \cite{Kolmogorov_etal1937} (FKPP) based on logistic-type growth, 
\citet{GertnerFreidlin1979} showed that the speed of the pulsating front that arises in such periodic flows can be obtained by solving an eigenvalue problem (see below). In practice, this procedure requires rather involved numerical computations; there is therefore  a need for simplified results that provide scaling predictions or closed-form expressions in asymptotic regimes. Results of this type have been derived in the limit of slow reactions and small diffusivity \cite{Audoly_etal2000,NovikovRyzhik2007}. Here we consider the opposite limit of fast reaction  (e.g. \cite{MajdaSouganidis1994}) relevant, for instance, to premixed flame propagation \cite{Peters}. In this limit, we obtain a compact expression for the speed  in terms of a single periodic path  that minimizes an action functional. This brings new physical insight, in particular into the role of the flow's stagnation points, and yields new closed-form results valid for a remarkably large range of reaction rates. 
It also provides  an efficient way to compute the speed in a regime where direct numerical computations are most challenging because of the widely disparate spatial scales  (see e.g. Fig.\ \fref{fronts}).

\begin{figure*}
\centerline{\includegraphics[scale=0.33]{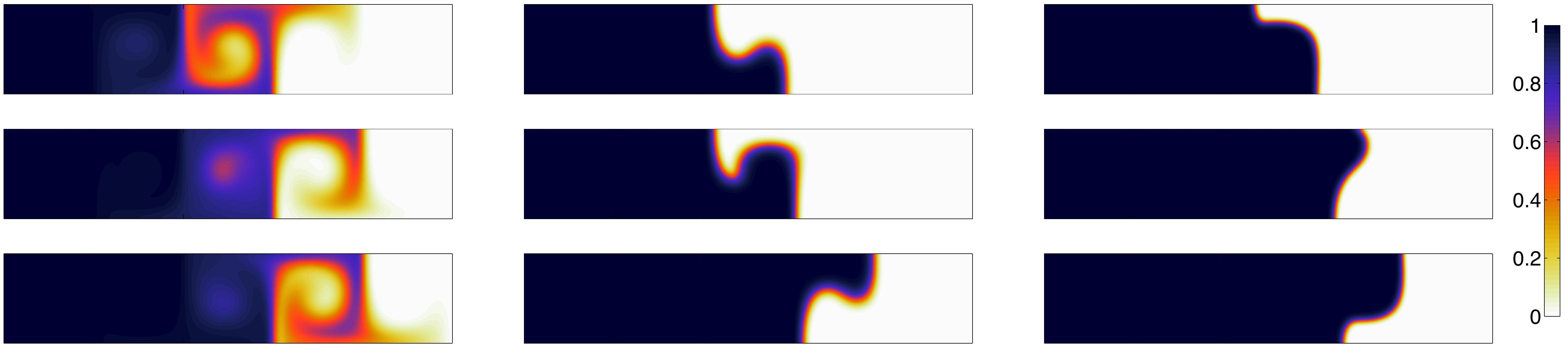}}
 \caption{
 (Color online). Successive snapshots of  the concentration $\theta$ for $\Pe=50$ and 
$\Da=0.5$  (left),
$\Da=5$  (middle), and 
$\Da=20$ (right) with time increasing from the top to the bottom rows. The corresponding front speeds are $c \approx 0.5$, $c \approx 1$, and $c \approx 1.6$.  
   }
 \flab{fronts}
 \end{figure*}

\textit{Model.}---The effect of a background  flow $\bu$ is incorporated in  the
FKPP 
 advection--diffusion--reaction equation
\beq\elab{FKPP}
 \partial_t\theta+\bu\cdot\nabla\theta=\Pe^{-1}\Delta\theta+
\Da\,r(\theta), 
\eeq 
 for the population concentration $\theta$. Here, the reaction term is $r(\theta)=\theta(1-\theta)$ or, more generally, any function $r(\theta)$ that 
satisfies 
$r(0)=r(1)=0$
with 
$r(\theta)>0$ for $\theta \in(0,1)$, $r(\theta)<0$ for $\theta \notin[0,1]$ and $r'(0)=\sup_{0<\theta<1} r(\theta)/\theta=1$.
The  non-dimensional parameters are   
the   P\'eclet and  Damk\"ohler numbers $\Pe=U\ell/\kappa$ and 
 $\Da=\ell/(U \tau)$ 
where $U$ and $\ell$ are the characteristic  amplitude and lengthscale of the flow, $\kappa$ the molecular diffusivity, and $\tau$  the reaction time. 
We consider the two-dimensional cellular flow $\bm{u}=(u_1,u_2)=(-\partial_y \psi,\partial_x\psi)$   
with streamfunction 
\beq\elab{psi}
\psi(x,y)=-\sin x\sin y.
\eeq
We take the domain to be the 
strip $D=(-\infty,\infty)\times[0,\pi]$ 
along which
an infinite array of identical cells are arranged,   
each of which  composed of two half-cells  
 of opposite circulation with hyperbolic stagnation points at each corner 
(streamlines are shown  for a single cell  in Fig.\ \fref{instantons}). 
As initial condition we take 
$\theta(x,y,0)=\Theta(-x)$,
where $\Theta$ is the Heaviside step function.
The boundary conditions are 
$\theta(\infty,y,t)= 0$ and 
$\theta(-\infty,y,t)= 1$, so that the front advances rightwards, and no flux, $\partial_y\theta= 0$ at $y=0, \, \pi$.
  The front characteristics change drastically with  $\Da$. When $\Da$ is small, the front's leading edge is confined near the cell boundaries (see left column in Fig.\ \fref{fronts}).  As $\Da$ increases, the front sharpens and the imprint of the flow, the boundary-layer structure in particular, is less  prominent  (see middle and right columns in Fig.\ \fref{fronts}).

The long-time speed of propagation of the front
is determined by the behavior of the solution near the front's leading edge,
where $0<\theta\ll 1$ and $r(\theta)\approx r'(0)\theta=\theta$. 
For steady periodic flows such as \eref{psi},  
this is given by  
\beq\elab{speed1}
 c=\inf_{q>0}\frac{f(q)+\Da}{q},
  \eeq
where $f(q)$ is the largest eigenvalue of
\begin{subequations}\elab{eval}
\beq
\mathcal{L}v=f(q)v,
 \eeq
with the  operator  $\mathcal{L}$ defined by
\beq
\mathcal{L}=\Pe^{-1}\Delta-\bu\cdot\nabla-2\Pe^{-1}q\partial_x+(u_1q+\Pe^{-1}q^2),
 \eeq
\end{subequations}
   and periodic and no-flux boundary conditions in $x$ and $y$, respectively
   [\citenum{GertnerFreidlin1979},\,\citenum{Freidlin1985},\,Ch.\ 7,\,\citenum{EvansSouganidis1989}].
%
 This result is intimately connected with the long-time 
large-deviation rate function associated with the concentration of the 
non-reacting passive scalar (i.e.,  $\Da=0$); specifically, $f(q)$ is the Legendre transform of 
this rate function
\cite{GertnerFreidlin1979,Freidlin1985,HaynesVanneste2014a,TzellaVanneste2014long}. 

In the absence of a background flow, 
$f(q)=q^2/\Pe$, recovering
the classical formula for the bare speed $c_0=2\sqrt{\Da/\Pe}=2\sqrt{\kappa/\tau}$ 
       (see Refs. \cite{Fisher1937,Kolmogorov_etal1937,LeachNeedham2003,vanSaarloos2003}  and references therein). 
For general $\bu\neq 0$, the eigenvalue problem (\ref{eqn:speed1})--(\ref{eqn:eval}) cannot be solved explicitly.

\textit{Small diffusivity, fast reaction.}---Our purpose  is to use asymptotic analysis to 
 obtain 
  the speed of the front in the large-P\'eclet limit with  
\[\gamma=\Da/\Pe=O(1),\]
corresponding to the geometric optics regime defined by $\kappa,\tau\to 0$ with $\kappa/\tau=O(1)$ \cite{Williams1985}.
This can be achieved by analyzing the large-$\Pe$ limit of the eigenvalue problem \eref{eval},
or by considering the large-$t$ limit of the geometric optics theory treatment of \cite[][Ch.\ 6]{Freidlin1985} (see also \cite{MajdaSouganidis1994, FreidlinSowers1999}).
It is however  convenient 
to start with the basic result exploited by 
 \cite{GertnerFreidlin1979,Freidlin1985}, namely that the front speed is controlled
by the point at which the solution to the linearization of equation \eref{FKPP} neither grows nor decays.

 We 
seek  a solution to this equation 
in the  WKB form 
\beq\elab{WKB}
\theta(x,y,t)\sim\exp(-\Pe(I(x,y,t)-\gamma t)), \quad\text{for $\Pe\gg 1$}, 
\eeq   
where $I(x,y,t)$ can be recognized as  the small-noise 
large-deviation rate function 
\cite{FreidlinWentzell1998}.
At leading-order, 
$\partial_tI+H(\nabla I,x,y)=0$,
where $H=\nor{\nabla I}^2+\bu(x,y)\cdot\nabla I$ 
 is the Hamiltonian and $\nor{\cdot}$  the usual  norm. 
 Its solution is well known from Hamilton--Jacobi theory (e.g.,
  \cite{Evans,FreidlinWentzell1998}) and  given by
\beq\elab{I}
	I(x,y,t)  =  \frac{1}{4}\inf_{\bphi(\cdot)}\int_0^t\nor{\dot{\bphi}(s)-\bu(\bphi(s))}^2ds, 
	\eeq
subject to  $\bphi(t)=(x,y)$ and $\bphi(0)=(0,\cdot)$. Thus,  
  	the behavior of \eref{WKB} is controlled by
	 a single path $\bphi^\ast(s)$ that
	minimizes 
     	this integral. This optimal path  is often called {\it instanton} \cite{Dykman_etal1994}.
    Expression \eref{WKB} then indicates that for $t\gg 1$,
the front speed satisfies
\beq\elab{gamma1}
\gamma=\lim_{t\to\infty}\frac{I(ct,y,t)}{t}\equiv\G(c),  
 \eeq
 where the limit $t \to \infty$ eliminates the dependence on $y$.  
This leads to 
\beq\elab{speed2}
c=\G^{-1}(\gamma).
\eeq

The front speed is therefore obtained by calculating $\G(c)$. 
  This  calculation is significantly simplified by observing that 
 	the  limit in  \eref{gamma1} is determined  in terms of periodic trajectories, an observation justified rigorously   in
	\cite{Piatnitski1998}. 
   	We take solutions  $\bphi(s)$ to be periodic 
in the sense that $\bphi(\tau)=\bphi(0)+(2\pi,0)$,
where the period is $\tau=2\pi/c$. 
    Letting $\sigma=c s$, we obtain the simplified expression 
		\beq\elab{G}
			\G(c)=\frac{1}{8\pi}\inf_{\bphi(\cdot)}\int_0^{2\pi}\nor{c{\bphi}'(\sigma)-\bu(\bphi(\sigma))}^2d\sigma, 
		\eeq
    subject to  $\bphi(2\pi)=\bphi(0)+(2\pi,0)$. 
		Expressions \eref{speed2} and \eref{G} are the main result of the paper. They provide a direct way of computing the instantons and thus the front speed.  
	 Note that $\G(c)$ may be interpreted as the Legendre transform of $\overline{H}$,
 the effective Hamiltonian 
  of the homogenized Hamilton--Jacobi equation
 $\partial_tI+\overline{H}(\nabla I)=0$ \cite{Lions_etal,MajdaSouganidis1994}. 
   In order to derive \eref{speed2}, we have formally assumed that $\gamma=O(1)$.  However,  the asymptotic results apply over a  broad range of values  of $\gamma$, specifically $\gamma \gg (\Pe \log \Pe)^{-1}$ as we show below.

\begin{figure}
 \centerline{\includegraphics[scale=0.37]{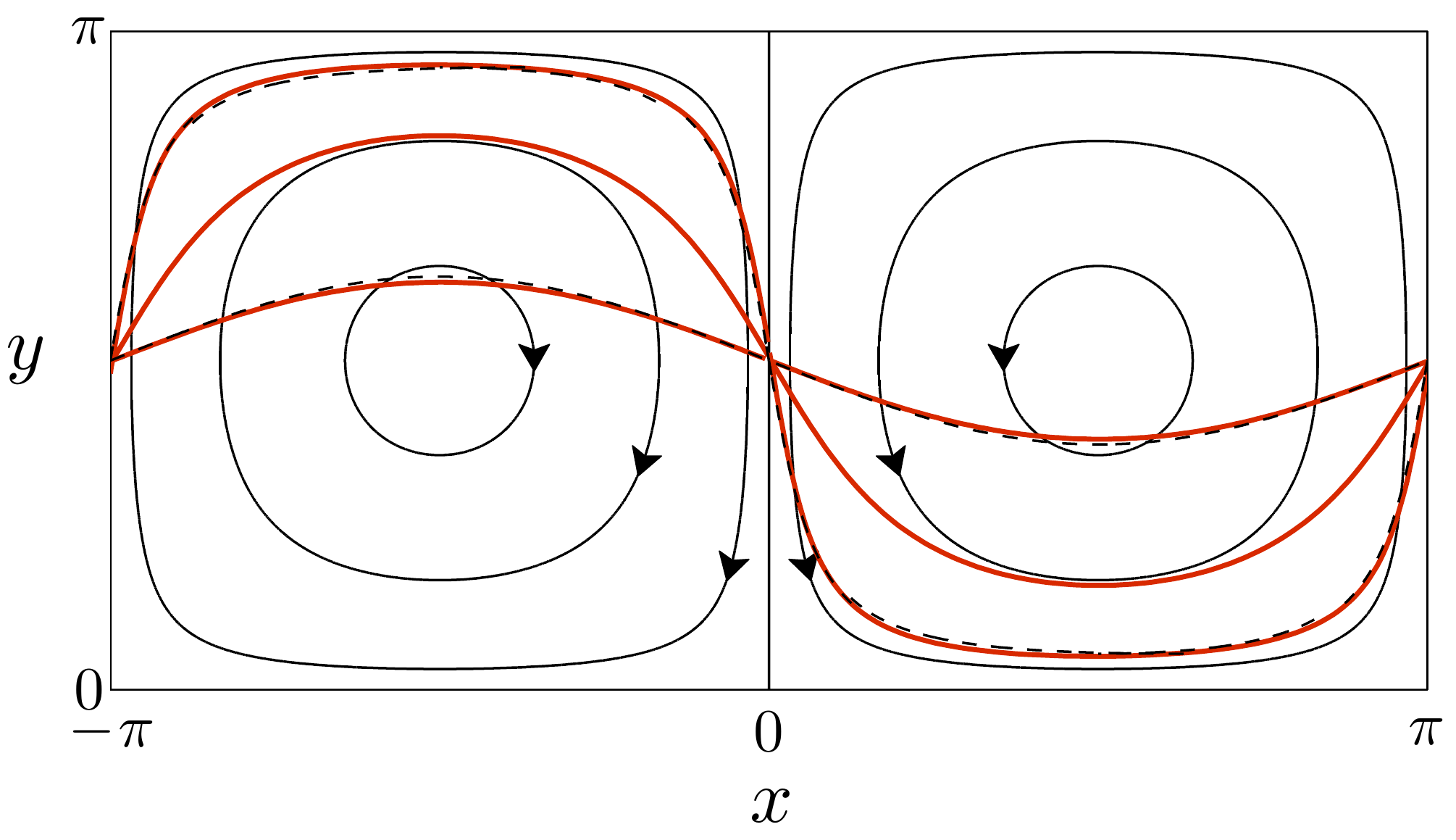}}
 \caption{
 (Color online). Streamlines of the cellular flow with streamfunction \eref{psi} (thin solid black lines) and trajectories of the instantons minimising \eref{G} (thick red lines). The instantons are calculated numerically  for $c=0.5$ (corresponding to $\gamma\approx 0.002$), $c=1$, ($\gamma\approx 0.09$) and $c=5$ ($\gamma\approx 5.9$) and become closer to the straight line $y=\pi/2$ as $c$ increases. The dashed lines show the small-$c$  and large-$c$ asymptotic approximations obtained using \eref{instantonsmallc} for $c=0.5$ and \eref{instantonlargec} for $c=5$, respectively.
   }
 \flab{instantons}
 \end{figure}

The solution to the minimization problem  \eref{G} is  easy to obtain numerically.	
We first discretize \eref{G} 
 and use MATLAB  to minimize the action. 
We iterate over $c$, starting with large values for which the straight line 
 $\bphi^\ast(s)=(cs,\pi/2)$  is a good initial guess. 
Characteristic examples of instantons obtained for different values of $c$ are shown in Fig.\ \fref{instantons} 
(the two cases with $c=0.5$ and $c=1$ correspond to the first two columns    in Fig.\ \fref{fronts}). 
For large values of $c$,  the instanton is close to a straight line. 
For small values of $c$, it follows  closely a streamline near the  cell boundaries.  
 Figure \fref{speed_asym} shows the behaviour of $c$ as a function of $\gamma$ deduced from \eref{speed2}.

\textit{Asymptotic limits.}---We now obtain closed-form expressions for the speed $c$ in two asymptotic limits. The first one corresponds to $\gamma \ll 1$ and hence $c \ll 1$. Numerical results (see  Fig.\ \fref{instantons} for $c=0.5$) suggest  that 
 the instanton 
 departs from the streamline only for  $y \approx \pi/2$ when it crosses the separatrix between adjacent cells.
 %
  It is also clear that \eref{G} is minimized when $\bphi^\ast(\sigma)=(x(\sigma),y(\sigma))$ satisfies $cy'\approx -\cos x\sin y$ (so that the instanton and flow speeds differ only in the $x$-direction).  
Exploiting symmetry to consider $0 \le \sigma \le \pi/2$ only, with $x(0)=0$, $y(0)=x(\pi/2)=\pi/2$ and $y'(\pi/2)=0$, we 
divide
%
the instanton  into two regions (see Fig.\ \fref{instantons}). In region $1$,  where
     $x\ll 1$, the integrand in \eref{G} is  approximately $(cx'-x\cos y)^2$, leading to the Euler--Lagrange equation $c^2 x''=x$ (since $cy'\approx-\sin y$). 
In   region $2$, $y\ll 1$, $cx'=\sin x\cos y\approx-\sin x$  and $cy'=-\cos x\sin y\approx-y\cos x$.  
Matching 
the solutions 
for $x,y\ll 1$ (the cell corner) gives the approximation
\beq\elab{instantonsmallc}
\bphi^\ast(\sigma)\sim
  \begin{cases}
   \left({C}_1(\sigma),{C}_2(\sigma)\right) & \text{for }\sigma \ll \pi/2 \\
    \left({C}_2(\pi/2-\sigma),{C}_3(\pi/2-\sigma)\right) & \text{for }\sigma \gg c
  \end{cases}
\eeq
where
\begin{align*}
C_1(\sigma)&=4 \exp\left({-\frac{\pi}{2c}}\right)\sinh\left(\frac{\sigma}{c}\right),\\
C_2(\sigma)&=2\tan^{-1}
\left(\exp\left(-\frac{\sigma}{c}\right)\right),
\\
C_3(\sigma)&=4 \exp\left(-\frac{\pi}{2c}\right)\cosh\left(\frac{\sigma}{c}\right).
   \end{align*}
Expression \eref{instantonsmallc} is in very good agreement with our numerical solution
(see Fig. \fref{instantons} for $c=0.5$).

\begin{figure}
\begin{center}
	\begin{overpic}[scale = .29]{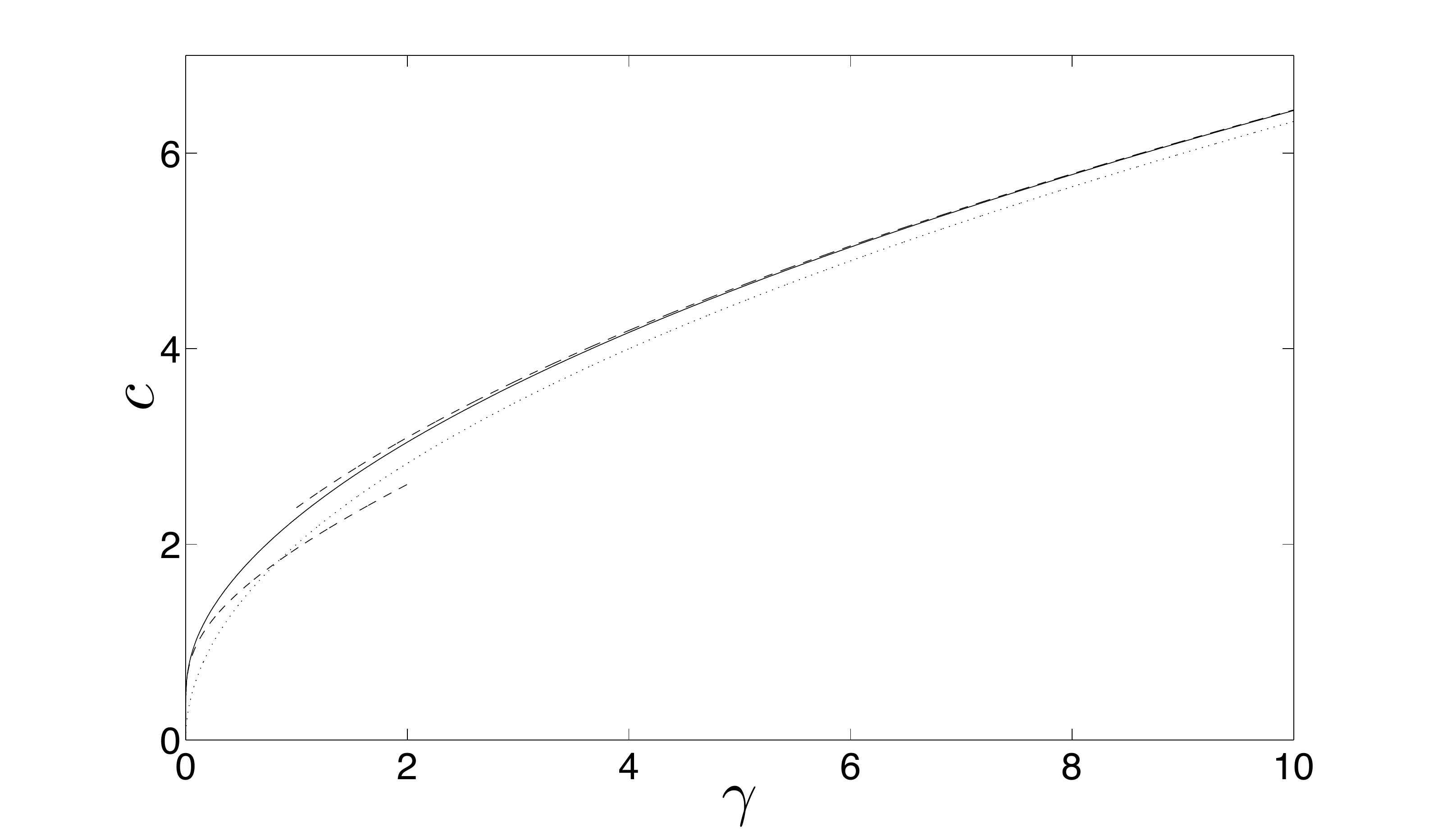}
		\put(52,9){\includegraphics[scale=0.15]{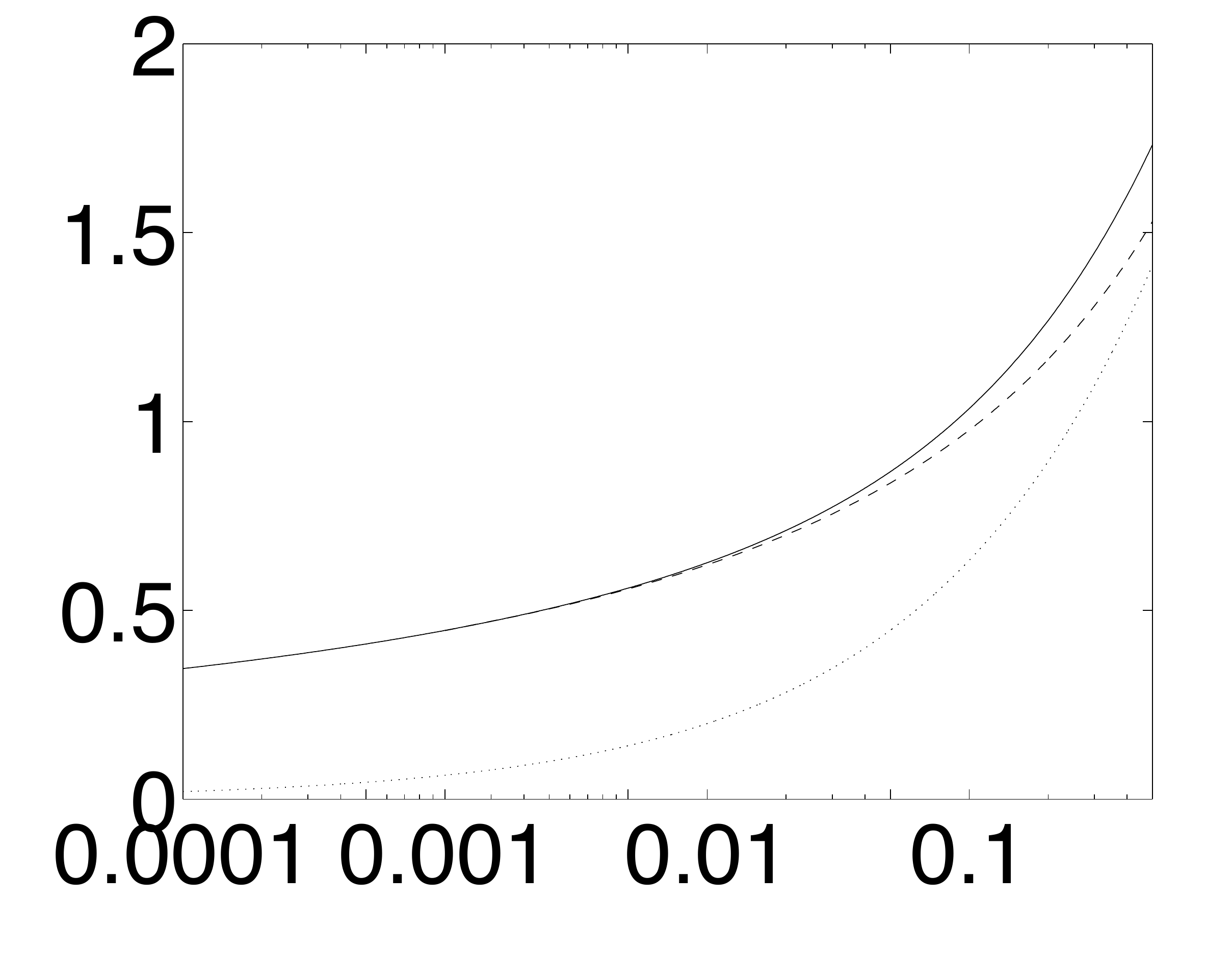}}
	\end{overpic}	
\end{center}
\caption{
Front speed $c$ for $\Pe \gg 1$ as a function of $\gamma$.
Prediction \eref{speed2} derived from the numerical minimization of \eref{G} (solid black line) is compared with 
its small-$c$ approximation \eref{c1app} (lower dashed line), its large-$c$ approximation \eref{c2app} (upper dashed line), and the bare speed $c_0=2\sqrt{\gamma}$ (dotted line).
The inset focuses on smaller values of $\gamma$. 
  }
 \flab{speed_asym}
 \end{figure}

Using \eref{instantonsmallc} gives the integrand in  \eref{G} as
$(cx'-x\cos y)^2\approx 16\exp\left(-\pi/c\right)\cosh^{-2}\left(\sigma/c\right)$, leading to
\beq\elab{G1}
\G(c)\sim  4\times(2/\pi)ce^{-\pi/c},\quad\text{where $c\ll 1$}
  \eeq
and the factor $4$ appears because, for  $\sigma \in [0\,\,2\pi]$, there are $4$ regions that are similar to region $1$.
 Inverting \eref{G1} and using \eref{speed2} finally gives   
\beq\elab{c1app} 
 c\sim\frac{\pi}{W_p(8\gamma^{-1})},\quad\text{for $\gamma\ll 1$,}
  \eeq
where $W_p$ denotes the principal real branch of the Lambert $W$ function
 \cite{NIST:DLMF}. This approximation
  holds for $(\log \Pe)^{-1} \ll \Da \ll \Pe$: 
  the lower bound follows from requiring that the argument of the exponential in \eref{WKB} be large, 
 $\Pe \,I\sim \Da\,\tau\gg 1$, where the period is roughly estimated as
 $\tau = 2 \pi/c \sim \log \Pe /2$ (see \eref{G1appb} below and \cite{TzellaVanneste2014long} for a complete argument).
Figure \fref{speed_asym} shows that 
this approximation  is excellent within its range of validity.
Since $\gamma \ll 1$, it is consistent to approximate
 $W_p(8\gamma^{-1})$ 
to  reduce \eref{c1app}  to
\beq\elab{G1appb}
c\approx\frac{\pi}{\log\Pe}.
\eeq
A qualitatively similar expression  was obtained in  
  \cite{Abel_etal2002,Cencini_etal2003}  using a heuristic approach  
 based on the so-called G-equation. 
To our knowledge, no equivalent expression to \eref{G1appb} has previously been derived from the FKPP equation 
  \eref{FKPP}. 
Note that the logarithmic dependence of the speed on $\Pe$ and its slow growth with $\Da$ (captured by \eref{c1app} 
but not \eref{G1appb}) is associated with the holdup of the instanton near the hyperbolic stagnation points at the cell corners.

The second limit leading to closed-form results corresponds to $\gamma \gg 1$, hence $c \gg 1$. In this case, we seek  an instanton  as a  power series 
$\bphi^\ast(\sigma)=(\sigma,y_0)+c^{-1}(x_1(\sigma),y_1(\sigma))+\ldots$,
where $x_1$, $y_1$ are functions of period $2\pi$ that satisfy $x_1(0)=y_1(0)=0$. 
Substituting  into \eref{G}, we find that at $O(c^{-1})$, 
\beq\elab{G2app1}
\G(c)=\frac{c^2}{4}+\frac{1}{8\pi}\inf_{x_1,y_0,y_1}
\int_0^{2\pi}
(x_1'^2+y_1'^2+4y_1\sin\sigma\sin y_0
 )
d\sigma,
   \eeq
 after some manipulations. 
Minimizing this integral leads to the instanton
\beq\elab{instantonlargec}
\bphi^\ast(\sigma)=\left(\sigma,\frac{\pi}{2}\right)+c^{-1}(0,-2\sin\sigma)+\ldots,
  \quad\text{for $c\gg 1$,}
\eeq
in excellent agreement with our numerical solution (see Fig.\ \fref{instantons} for $c=5$).
Combining \eref{G2app1}
and \eref{instantonlargec}, yields  
$\G(c)=c^2/4-3/8+O(c^{-2})$. We now use
 \eref{speed2} to find that 
\beq\elab{c2app} 
c
\sim 2\sqrt{\gamma}\left(1+\frac{3}{16\gamma}+\ldots\right),
\quad\text{for $\gamma\gg 1$,} 
\eeq
which corresponds to $\Da \gg \Pe$. 
The leading-order term in \eref{c2app} is  the bare speed $c_0$. As Fig.\ \fref{speed_asym} shows, the second term in the expansion is necessary for a good agreement between  asymptotic and  full results.

\begin{figure}
     \centerline{\includegraphics[scale=0.29]{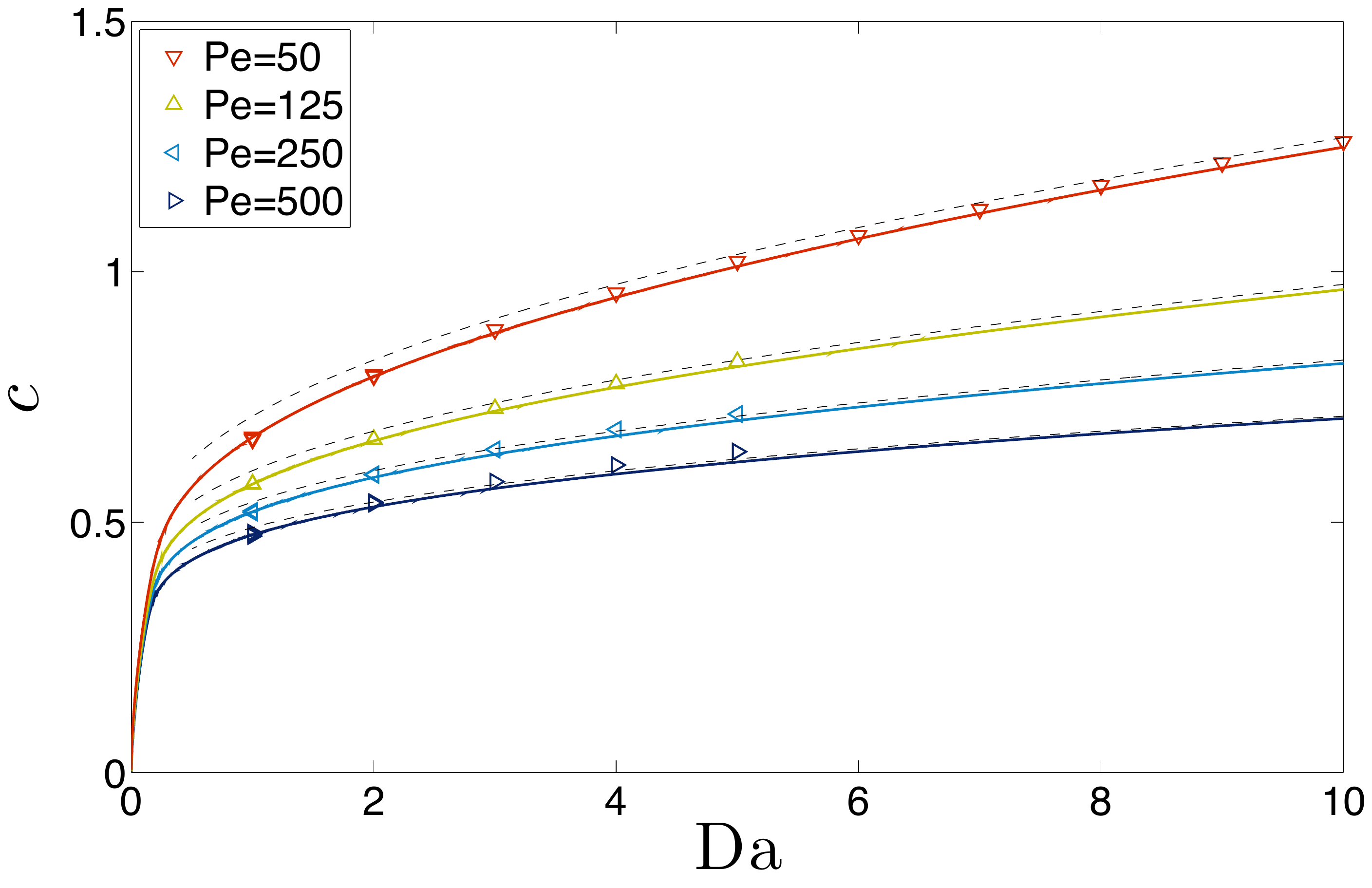}}
 \caption{
 (Color online).  Front speed $c$ as a function of $\Da$ for different values of $\Pe$.
 The large-$\Pe$ prediction \eref{speed2} derived from the numerical minimization of \eref{G} (dashed lines) is compared with the exact expression  \eref{speed1} estimated by solving the eigenvalue problem \eref{eval} numerically (solid lines) and with direct   numerical simulations of \eref{FKPP} (symbols). 
      }
 \flab{speed}
 \end{figure}

\textit{Comparison with numerical results.} We now compare our predictions for $c$  derived from \eref{speed2}--\eref{G}  
with  values obtained from (i) numerical evaluation of the principal eigenvalue 
 in \eref{eval}, and (ii)  direct numerical simulations of \eref{FKPP} with $r(\theta)=\theta(1-\theta)$.
For (i) we use  a standard second-order 
discretization to approximate  \eref{eval} and choose the spatial resolution $\Delta$ to 
satisfy $\pi/\Delta=750$.
The resulting matrix eigenvalue problem is solved for a range of values of $q$ using  
MATLAB. 
For (ii) we discretize
\eref{FKPP} using a fractional-step method with a Godunov splitting 
 which  alternates between advection (using a first-order upwind method with a minmod limiter -- see \cite{Leveque} for  details), diffusion (using an alternating direction implicit method) and reaction (solved exactly). 
%
%
We choose the same spatial resolution $\Delta$ as for 
\eref{eval}. 
The computational domain is made finite using artificial boundaries at $x=\pm N \pi$, with 
$N=5$, so that boundary effects are negligible. 
The front is tracked for long times by modifying the computational domain: 
when the solution  at $x=(N-1)\pi$  exceeds 
$\delta=10^{-6}$, we eliminate the nodes with
$-N\pi\leqslant x \leqslant (-N+1)\pi$
and add new nodes with 
$N\pi\leqslant x \leqslant (N+1)\pi$ 
where we set $\theta= 0$.
%
%
%
 We calculate the front speed using a linear fit of the right endpoint of the front,  $x^{+}_\epsilon(t)=\text{max}\{x:\theta(x,t)=\epsilon\}$ 
where $\epsilon=10^{-3}$. 
%
%
Results are insensitive to the exact values of $\epsilon$ and $\delta$. 

The three set of numerical  results are shown in Fig. \fref{speed}.
The speeds derived from the eigenvalue equation \eref{eval} are in excellent agreement with the corresponding values obtained from the full numerical simulations of equation \eref{FKPP}. 
 With increasing values of $\Pe$,   the asymptotic expression \eref{speed2}--\eref{G}  becomes increasingly accurate, with excellent agreement for $\Pe=250,500$ and satisfactory agreement for the moderate values $\Pe=50,125$.
As expected, \eref{speed2}--\eref{G} is valid for a broad range of values of $\Da$, restricted only by $\Da\gg(\log\Pe)^{-1}$.  
Note that  the use of both the eigenvalue equation and the full numerical simulations  is restricted: as $\Pe$ increases, 
 the solutions to \eref{FKPP} and \eref{eval} become progressively localized, with $O(1/\sqrt{\Da\Pe}$) lengthscales that are challenging  to resolve when $\Da, \, \Pe \gg 1$.

\textit{Conclusion.}---We have derived a compact expression 
for the front speed 
 %
 based on the minimization of the large-deviation action \eref{G} over   periodic instantons. 
This leads to the efficient computation of the speed 
for a large range of values of $\Da$.   
For the particular case of cellular  flows, 
this expression provides the new closed-form results 
\eref{c1app} and \eref{c2app} valid for $(\log \Pe)^{-1} \ll \Da\ll\Pe$ and $\Da\gg\Pe$. 
In the first regime, the passage of the front near the stagnation points at the cell corners is shown to control the front speed; as a result this is almost insensitive to the reaction rate and depends logarithmically on the P\'eclet number.
For $\Da = O(\log \Pe)^{-1}$ and smaller, the front speed is not controlled by a single minimizing trajectory, and asymptotic solutions to the eigenvalue problem \eref{eval} must be sought by other means; this will be the subject of future work \citep{TzellaVanneste2014long}.

\smallskip

The authors thank P. H. Haynes, G. C. Papanicolaou and A. Pocheau for  helpful discussions. 
The work was  supported by EPSRC (Grant No.\ EP/I028072/1).

\bibliography{front}

  \end{document}